\documentclass[seceq,supplement]{ptptex}

\title{
Generalized Green-Kubo formula for a dissipative quantum system%
}

\author{
Hisao \textsc{Hayakawa}%
}

\inst{
Yukawa Institute for Theoretical Physics, Kyoto University,  Kitashirakawaoiwake-cho, Sakyo-ku,
Kyoto 606-8502, Japan
}

\abst{
A generalized Green-Kubo formula is derived for a quantum dissipative system of driven Brownian particle, 
in which the coupling between the system and the environment is linear. 
The structure is essentially the same as that for the generalized Green-Kubo formula for driven granular particles.
It is demonstrated that the correction to the conventional Green-Kubo formula is zero for
 a free Brownian particle.
}

\begin{document}

\maketitle

\section{Introduction}

Green-Kubo formula\cite{kubo} is one of the most fundamental relations in nonequilibrium statistical physics. 
The original derivation is restricted to the linear nonequilibrium case,
but number of generalizations are proposed by many researchers\cite{evans,zubarev,saito07,fujii,shimizu}, 
though the relationship 
among their formulations is not still well understood.

We believe that roles of dissipation in Green-Kubo formula should be clarified, though
it is unclear in the original derivation.
Indeed, the Green-Kubo formula defines the transport coefficient which represents the dissipation.
Moreover, if there is no dissipation in a system, the time integral of current correlation function in the formula 
 should diverge.
Therefore, purely mechanical derivation of Green-Kubo formula might be misleading, but correct derivation should
include the dissipation explicitly.

Recently, Chong et al.\cite{chong} have derived a new generalized Green-Kubo formula 
for driven dissipative and  classical particles 
as a natural extension of that by Evans and Morriss.\cite{evans} Their derivation has several remarkable points; 
(i) the formulation can include the integral fluctuation theorem without microscopic time-reversal symmetry,
(ii) the role of dissipation is clear in their derivation, and
(iii) one can develop the nonequilibrium mode-coupling theory for sheared granular liquids or sheared glassy systems.
Since their derivation is so general that one can expect that their method can be used for quantum cases.
In this paper we demonstrate how to apply such a formula to quantum systems.

In this paper, we focus on a nonequilibrium steady state of a quantum Brownian partilce. 
This is because (i) the simplest system among open quantum systems\cite{breuer,weiss,CL83,Hu,saito00}, 
(ii)  we know the origin of dissipation of quatum Brownian system as the energy flux between the system and
the environment, and (iii) there is the quantum version of the violation of fluctuation-dissipation relation.\cite{saito08}   

The organization of this paper is as follows. In the next section, we specify the basic equations to be analyzed in this paper.
In section 3, we will obtain an exact solution of our model. 
In section 4, we will present the formal form of generalized Green-Kubo formula whose bilinear form is reduced to the conventional
Green-Kubo formula. 
In section 5, we will focus on the case that a Brownian particle in a harmonic potential, in which the integration 
involved in the generalized Green-Kubo formula can be carried out.
In section 6, we will discuss and conclude our results.   
We also invole two Appendices, where Appendix A is devoted to the derivation of generalized Kubo's identity, and
Appendix B gives an explicit calculation of the time evolution of the momentum of the Brownian particle.

\section{Model}

Let us begin with a quantum master equation for a Brownian particle\cite{breuer} under an external field $F_{\rm ex}$,
 which is the essentially same as Caldeira-Leggett model\cite{CL83}.
Note that
the generalization to $N$ Brownian particles is straightforward from  that presented here. For simplicity, we restrict our interest to
a particle coupled with the heat bath.

Let us consider a Brownian particle of mass $m$ with its coordinate $x$ and momentum $p$ in a potential $V(x)$
 under a steady external field $F_{ex}$.\footnote{Of course, the external field can be involved in the potential, 
but we separate the contribution of the external field from the stationary potential $V(x)$. }
The particle is assumed to be coupled with a bath consisting of a large number of harmonic oscillators with masses $m_n$ and
frequencies $\omega_n$. Thus, the  total Hamiltonian might be written as
\begin{eqnarray}
H &=& H_S+H_C+H_B+H_I+H_{\rm ex} \nonumber\\
&=& \frac{p^2}{2m}+V(x)- xF_{\rm ex} +\sum_n \left\{\frac{p_n^2}{2m_n}+\frac{1}{2}m_n\omega_n^2\left(x_n-\kappa_n \frac{x}{m_n\omega_n^2}
\right)^2\right\}
\end{eqnarray}
Here, 
the system Hamiltonian of the particle is given by
\begin{equation}\label{3.375}
H_S=\frac{p^2}{2m}+V(x),
\end{equation}
where we do not specify the form of the potential in the formulation in the main part of our paper.
The bath Hamiltonian is represented by
\begin{equation}\label{3.376}
H_B=\sum_n \hbar \omega_n \left(b_n^{\dagger}b_n+\frac{1}{2}\right)=
\sum_n\left(\frac{1}{2m_n}p_n^2+\frac{1}{2}m_n\omega_n^2x_n^2\right).
\end{equation}
Here, $b_n^{\dagger}$ and $b_n$ denote the Bosonic creation and the annihilation operators of the bath, respectively,
while $x_n$ and $p_n$ are the corresponding coordinate and the momentum. 
Similarly, the interaction Hamiltonian is given by
\begin{equation}\label{3.377}
H_I=-x \sum_n \kappa_n x_n = -x B ,
\end{equation}
where $\kappa_n$ is the coupling constant, the bath operator is
\begin{equation}\label{3.378}
B\equiv \sum_n \kappa_n x_n =\sum_n \kappa_n \displaystyle\sqrt{\frac{\hbar}{2m_n\omega_n}}(b_n+b_n^{\dagger}) .
\end{equation}
We introduce the external Hamiltonian coupled with the external force $F_{\rm ex}$ as
\begin{equation}\label{H_ex}
H_{\rm ex}=-x F_{\rm ex}.
\end{equation}
Note that $H_{\rm ex}$ can be absorbed in $H_{S}$ but we separate its contribution to clarify the response to the external field.
We also introduce the counter-term Hamiltonian:
\begin{equation}
H_C= {\cal K} x^2 \equiv
x^2\sum_n \frac{\kappa_n^2}{2m_n \omega_n^2} .
\end{equation} 
This counter-term Hamiltonian can be absorbed in the potential term $V(x)$ as $V_{\rm eff}(x)=V(x)+H_C$. 
However, if we regard the unperturbed Hamiltonian as $H_S+H_B$ without couping between the system and the bath at $t=0$, 
the effect of $H_C$ appears in later expressions. 
We also note that $H_C$ must be treated as a term second order in the coupling,
while $H_I$ is of the first order.

Caldeira and Leggett\cite{CL83} were interested in the low frequency bahavior of particles. Then they adopted the simple assumption
 $\hbar\omega_0\ll {\rm Min}\{ \hbar\Omega, 2\pi kT \}$ under the condition
$V(x)\approx (1/2)m\omega_0^2x^2+O(x^3)$  with
a high frequency cutoff $\Omega$, the Boltzmann constant $k$ and the temperature $T$:
\begin{equation}\label{B3.398}
x_S(-\tau) \equiv e^{-i H_S\tau/\hbar} x e^{i H_S \tau/\hbar} \approx x-\frac{i}{\hbar}[H_S,x] \tau .
\end{equation}
Here, however, we do not have to use this quasi-classical expression (\ref{B3.398}) 
for our argument. Here, we have introduced the commutation relation $[A,B]\equiv AB-BA$.

The starting equation is the Born-Markov approximation for the reduced density matrix of the Brownian particle,
which obeys\cite{breuer}
\begin{equation}\label{B3.382}
\frac{d}{dt}\rho_S(t)=-\frac{i}{\hbar}[H_S+H_C,\rho_S(t)]+\frac{i F_{\rm ex}}{\hbar}[x,\rho_S(t)] 
-\frac{1}{\hbar}\int_0^{\infty}d\tau 
{\rm tr}_B[H_I,[H_I(-\tau),\rho_S(t)\otimes \rho_B]] .
\end{equation}
Hereafter, we adopt the interaction picture with the respect to the unperturbed Hamiltonian $H_0=H_S+H_B$.
We shall assume that the initial condition satisfies 
\begin{equation}\label{initial_rho}
\rho_S(0)=\rho_{\rm eq}\otimes \rho_B ,
\end{equation}
where
\begin{equation}
\rho_{\rm eq}=\frac{\exp[-\beta H_S]}{{\rm tr}_S\exp[-\beta H_S]}, \qquad
\rho_B=\frac{\exp[-\beta H_B]}{{\rm tr}_B \exp[-\beta H_B]} \quad 
\end{equation}
with $\beta=1/kT$.
Note that the assumption on $\rho_S(0)$ might be removable. 
Indeed, the steady distribution $\rho_S(t\to \infty)$ of Caldeira-Leggett model with Eq. (\ref{B3.398}) is relaxed
to $\rho_S(0)$\cite{breuer,Hu}, if the particle is trapped in a potential.

For the discussion of quantum Brownian motion we introduce the spectral functions
\begin{equation}\label{B3.385-386}
D(\tau)\equiv i \langle [B, B_B(-\tau) ] \rangle_B, \qquad
D_1(\tau) \equiv \langle \{B, B_B(-\tau) \} \rangle_B ,
\end{equation}
where $B_B(t)\equiv e^{iH_Bt/\hbar}B e^{-i H_Bt/\hbar}$ and $\{A,B\}\equiv AB+BA$.
 Note that $H_S$ and $H_B$ are decoupled with each other 
in the unperturbed state, where $\langle \cdots \rangle_B$ represents the average in terms of the density matrix $\rho_B$.
We also note that $D(\tau)$ and $D_1(\tau)$ are respectively referred to the dissipation and the noise kernel.
Making use of the spectral density
\begin{equation}
J(\omega)=\sum_n \frac{\kappa_n^2}{2m_n\omega_n}\delta(\omega-\omega_n)  ,
\label{B3.387}
\end{equation}
we can write the explicit representations for the correlation functions
\begin{eqnarray}
D(\tau) &=& 2\hbar \int_0^{\infty}d\omega J(\omega) \sin\omega \tau  , \\
D_1(\tau) &=& 2\hbar \int_0^{\infty}d\omega J(\omega) 
{\rm coth}
\left(
\frac{\beta\hbar\omega}{2}
\right)
 \cos\omega \tau  .
\end{eqnarray}
After straightforward calculation, eq.(\ref{B3.382}) can be rewritten as\cite{breuer}
\begin{eqnarray}\label{B3.390}
& &\frac{d}{dt}\rho_S(t)
=-\frac{i}{\hbar}[H_S+H_C,\rho_S(t)]+\frac{i F_{\rm ex}}{\hbar}[x,\rho_S(t)] 
\nonumber\\
& &\qquad +\frac{1}{2\hbar^2}\int_0^{\infty}d\tau
\left( i D(\tau)[x,\{x_S(-\tau),\rho_S(t)\}]-D_1(\tau)[x,[x_S(-\tau),\rho_S(t)]]\right),
\end{eqnarray}
which is the basic equation of this paper. 

The properties of the second line of (\ref{B3.390}) strongly depend on the behaviour of the dissipation and the noise 
which are determined by $J(\omega)$. We adopt  a continous distribution of the bath modes and replace the spectral
density by a smooth function of $\omega$ for the explicit calculation. 

For the calculation in section 5, we should specify the form of $J(\omega)$.
Here we adopt the Ohmic spectral density with Lorentz-Drude cutoff function, $J(\omega)$ is given by
\begin{equation}\label{J(omega)}
J(\omega)=\frac{2m\gamma}{\pi}\omega \frac{\Omega^2}{\omega^2+\Omega^2} ,
\end{equation}
where $\gamma$ is a damping constant and $\Omega$ is a high frequency cutoff. Note that $J(\omega)$ satisfies Ohmic dispersion
$J(\omega)\to 2m\gamma \omega/\pi$ as $\omega\to 0$.
In this case, $D(\tau)$ and $D_1(\tau)$ are respectively given by
\begin{equation}\label{D(t)_Drude}
D(\tau)=2m\gamma \hbar \Omega^2 e^{-\Omega \tau}
\end{equation}
and
\begin{equation}
\label{D_1(t)_Drude}
D_1(\tau)=4m\gamma k T \Omega^2\sum_{n=-\infty}^{\infty}\frac{\Omega e^{-\Omega \tau}-|\nu_n|e^{-|\nu_n|\tau}}{\Omega^2-\nu_n^2}
\end{equation}
for $\tau>0$, where
$\nu_n\equiv 2\pi n kT/\hbar$ is known as the Matsubara frequency.\cite{breuer}

\section{The solution of Liouville equation}

Let us rewrite Eq.(\ref{B3.390}) as
\begin{equation}\label{L1}
\frac{d}{dt}\rho_S(t)=-i {\cal L}^{\dagger} \rho_S(t) ,
\end{equation}
where the Liouville operator is given by
\begin{equation}\label{L2}
i{\cal L}^{\dagger}\equiv \frac{i}{\hbar}[H_S+H_C,]-\frac{i F_{\rm ex}}{\hbar}[x,]
-\frac{1}{2\hbar^2}\int_0^{\infty}d\tau (i D(\tau)[x,\{x_S(-\tau),\}]-D_1(\tau)[x,[x_S(-\tau),]]) .
\end{equation}
Since this Liouville operator $i{\cal L}^{\dagger}$ is independent of time, we can use the identity
\begin{equation}\label{L3}
e^{-i{\cal L}^{\dagger}t}=1+\int_0^t dse^{-i {\cal L}^{\dagger}s}(-i {\cal L}^{\dagger}) .
\end{equation}
Substituting Eq.(\ref{initial_rho}) into Eq. (\ref{L3}) we obtain
\begin{equation}
\rho_S(t)=\rho_{\rm eq}+\int_0^t ds e^{-i {\cal L}^{\dagger}s}(-i {\cal L}^{\dagger})\rho_{\rm eq} .
\label{L6}
\end{equation}
Here, $-i {\cal L}^{\dagger}\rho_{\rm eq}$ consists of four terms:
\begin{eqnarray}\label{L8}
-i {\cal L}^{\dagger}\rho_{\rm eq}
&=&\frac{i F_{\rm ex}}{\hbar}[x, \rho_{\rm eq}]-\frac{i}{\hbar}{\cal K}[x^2,\rho_{\rm eq}]
+\frac{i}{2\hbar^2}\int_0^{\infty}d\tau D(\tau)[x,\{x_S(-\tau),\rho_{\rm eq}\}]
\nonumber\\
& &-\frac{1}{2\hbar^2}\int_0^{\infty}d\tau D_1(\tau)[x,[x_S(-\tau),\rho_{\rm eq}]] .
\end{eqnarray}

The first term on the right hand side of eq.(\ref{L8}) produces the conventional Green-Kubo formula.
With the aid of Kubo's identity (\ref{kubo1}), we can rewrite
\begin{eqnarray}
[x,\rho_{\rm eq}]&=&
\rho_{\rm eq}\int_0^{\beta}d\lambda e^{\lambda H_S}[H_S,x]e^{-\lambda H_S} \nonumber\\
&=&-\frac{i\hbar}{m} \rho_{\rm eq} \int_0^{\beta}d\lambda p_S(-i\hbar \lambda) ,
\end{eqnarray}
where we have used $[H_S,x]= -i\hbar p/m$.
Thus, the first term on the right hand side of (\ref{L8}) is reduced to
\begin{equation}\label{L8-1}
\frac{i F_{\rm ex}}{\hbar}[x,\rho_{\rm eq}]
=\frac{F_{\rm ex}}{m} \rho_{\rm eq}\int_0^{\beta}d\lambda p_S(-i\hbar \lambda) .
\end{equation}

Similarly, the second term on the right hand side of (\ref{L8}) which is the order of square of the coupling constant 
can be calculated. From (\ref{kubo1}) we readily obtain 
\begin{equation}
[x^2,\rho_{\rm eq}]=-\frac{i \hbar \rho_{\rm eq}}{m} \int_0^{\beta}d\lambda
 \{x_S(-i\hbar \lambda), p_S(-i\hbar \lambda) \} ,
\end{equation}
where we have used $[H_S,x^2]=-i\hbar (xp+px)/m$.
Thus, the second term on the right hand side of (\ref{L8}) is reduced to
\begin{equation}
\label{L8-2}
-\frac{i}{\hbar}{\cal K}[x^2,\rho_{\rm eq}]
=-\frac{{\cal K} \rho_{\rm eq}}{m}
\int_0^{\beta}d\lambda \{x_S(-i\hbar \lambda), p_S(-i\hbar \lambda) \} .
\end{equation}

The contributions from the third term and the fourth term on the right hand side of (\ref{L8}) are more complicated.
From  (\ref{kubo1}), (\ref{kubo2}) and (\ref{kubo3}) we obtain an identity
\begin{eqnarray}
[x, \{x_S(-\tau),\rho_{\rm eq}\}]&=&
\rho_{\rm eq} \left( [x,x_S(-\tau)]+[x,x_S(-i\hbar \beta-\tau)] \right) \nonumber\\
& &
-\frac{i\hbar}{m} \rho_{\rm eq} \int_0^{\beta}d\lambda p_S(-i\hbar\lambda)(x_S(-\tau)+x_S(-i\hbar \beta-\tau)) .
\end{eqnarray}
Similarly, thanks to (\ref{kubo1}) we obtain
\begin{eqnarray}
[x, [x_S(-\tau), \rho_{\rm eq}]]
&=&
 \rho_{\rm eq} \left([x, x_S(-i\hbar \beta-\tau)]-[x,x_S(-\tau)] \right)\nonumber\\
& &
-i\hbar \frac{\rho_{\rm eq}}{m}\int_0^{\beta}d\lambda p_S(-i\hbar \lambda) (x_S(-i\hbar\beta-\tau)-x_S(-\tau))
.
\end{eqnarray}
Thus, the third term and the fourth term on the right hand side of (\ref{L8}) is reduced to
\begin{eqnarray}
& & 
\frac{i}{2\hbar^2}\int_0^{\infty}d\tau D_1(\tau)[x,\{x_S(-\tau),\rho_{\rm eq}\}]
-\frac{1}{2\hbar^2}\int_0^{\infty}d\tau D(\tau)[x,[x_S(-\tau),\rho_{\rm eq}]]
\nonumber\\
& & =
\frac{\rho_{\rm eq}}{2\hbar^2}\int_0^{\infty}d\tau \tilde{D}_+(\tau)\{
[x,x_S(-\tau)]-\frac{i\hbar}{m} \int_0^{\beta}d\lambda p_S(-i\hbar\lambda)x_S(-\tau)
\}
\nonumber\\
& & {~~}
-\frac{\rho_{\rm eq}}{2\hbar^2}\int_0^{\infty}d\tau\tilde{D}_-(\tau)\{[x,x_S(-\tilde{\tau})]
-\frac{i\hbar}{m} \int_0^{\beta}d\lambda p_S(-i\hbar\lambda)x_S(-\tilde{\tau})\}
 ,
\end{eqnarray}
where $\tilde{\tau}=\tau+i\hbar\beta$ and
\begin{equation}
\tilde{D}_{\pm}(\tau)\equiv 
D_1(\tau)\pm i D(\tau)
.
\end{equation}
Thus, we can write
\begin{equation}\label{rho_S(t)}
\rho_S(t)=\rho_{\rm eq}+\int_0^tds e^{-i{\cal L}^{\dagger}s}[\rho_{\rm eq} \Theta],
\end{equation}
where 
\begin{eqnarray}\label{Omega}
\Theta &\equiv&
 \frac{F_{\rm ex}}{m}\int_0^{\beta}d\lambda p_S(-i\hbar\lambda)
 -\frac{{\cal K}}{m} \int_0^{\beta}d\lambda \{x_S(-i\hbar\lambda), p_S(-i\hbar \lambda)\} 
 \nonumber\\
 & &+\frac{i}{2\hbar m}\int_0^{\infty}d\tau 
\LARGE\{\tilde{D}_-(\tau)\int_0^{\beta}d\lambda p_S(-i\hbar\lambda)x_S(-\tau-i\hbar\beta) \nonumber\\
& & {~}
-\tilde{D}_+(\tau) \int_0^{\beta}d\lambda p_S(-i\hbar\lambda)x_S(-\tau)\LARGE\}
\nonumber\\
& &+ {~}\frac{1}{2\hbar^2}\int_0^{\infty}d\tau 
\left( \tilde{D}_+(\tau)[x,x_S(-\tau)] 
-\tilde{D}_-(\tau)[x,x_S(-\tau-i\hbar\beta)]
\right) .
\end{eqnarray}

Before closing this section, we should note an important property of $\Theta$ which satisfies
\begin{equation}\label{<Theta>=0}
\langle \Theta \rangle_{\rm eq}\equiv {\rm tr}_S\{\rho_{\rm eq} \Theta \}=0 .
\end{equation}
This relation is easily verified from Eq.(\ref{L8}) with the invariant property of the trace under a cyclic permutation.

\section{Generalized Green-Kubo formula}

Let us derive the generalized Green-Kubo formula.
For simplicity, we discuss the average behavior of the momentum $p$
\begin{equation}
\langle p \rangle_t\equiv {\rm tr}_S\{ \rho_S(t)p \}
\label{def_<p>} .
\end{equation}
Substituting (\ref{def_<p>}) into (\ref{rho_S(t)}) we obtain
\begin{equation}
\langle p\rangle_t={\rm tr}_S \{\rho_{\rm eq} p\}+\int_0^t ds {\rm tr}_S\{e^{-i {\cal L}^{\dagger} s}
[\rho_{\rm eq}\Theta] p\}  .
\end{equation}
With the help of the property 
\begin{equation}
{\rm tr}_S\{e^{-i {\cal L}^{\dagger} s}[\rho_{\rm eq}\Theta] p\}
={\rm tr}_S\{\rho_{\rm eq} \Theta p_H(t) \}
\end{equation} 
with $p_H(t)\equiv e^{i{\cal L}t}p=e^{iHt/\hbar}pe^{-iHt/\hbar}$, we can write the generalized Green-Kubo formula,
where the Liouville operator $i{\cal L}$ satisfies Heisenberg's equation of motion
$\dot{A}_H(t)=i{\cal L} A_H(t)=\frac{i}{\hbar}[H,A_H(t)]$. 
We note that the last equality for $p_H(t)$ holds because of our special set-up, where we adopt the basic model under
the Born-Markovian approximation with the linear couping between the system and the environment.

The explicit form of $i {\cal L}$ is given by
\begin{eqnarray}\label{iL}
i{\cal L}&=&\frac{i}{\hbar}[H_S+H_C,]-\frac{i F_{\rm ex}}{\hbar}[x,] \nonumber\\
& &-\frac{1}{2\hbar^2}\int_0^{\infty}d\tau\left(
iD(\tau)\{x_S(-\tau),[x,]\}+D_1(\tau)[x_S(-\tau),[x,]]\right),
\end{eqnarray}
where we have used the invariant property of the trace under a cyclic permutation.

We can itroduce
\begin{equation}\label{phase-shrink}
\Lambda\equiv i{\cal L}^{\dagger}-i{\cal L} ,
\end{equation} 
whose expectation value corresponds to the phase volume contraction in classical situations.
The operator $\Lambda$ is immediately obtained as
\begin{eqnarray}\label{Lambda}
\Lambda&=&\frac{1}{2\hbar^2}\int_0^{\infty}d\tau D_1(\tau)\left([x,[x_S(-\tau),]]+[x_S(-\tau),[x,]]\right)
\nonumber\\ &&
 -\frac{i}{2\hbar^2}\int_0^{\infty}d\tau D(\tau)\{[x,x_S(-\tau)],  \}.
\end{eqnarray} 
Thus, we expect $\langle \Lambda \rangle_t$ represents the quantum counter-part of the phase volume contraction.

In the steady state limit, we should take the limit of $t\to \infty$ as 
\begin{equation}
\langle p \rangle_{\rm SS}\equiv \lim_{t\to\infty}\langle p \rangle_t
=\int_0^{\infty}dt {\rm tr}_S[\rho_{\rm eq} \Theta p_H(t)] .
\end{equation} 

The contribution of the first term on the right hand side of (\ref{Omega}) is
\begin{equation}
\langle p \rangle_{\rm SS}^{(1)}
=\frac{F_{\rm ex}}{m}\int_0^{\infty}dt \int_0^{\beta}d\lambda\langle p_S(-i\hbar \lambda) p_H(t) \rangle_{\rm eq} ,
\label{GK1}
\end{equation} 
which is the conventional Green-Kubo formula.
The contribution of the second term on the right hand side of (\ref{Omega}) is
\begin{equation}\label{GK2}
\langle p \rangle_{\rm SS}^{(2)}
=-\frac{{\cal K}}{m} \int_0^{\infty}dt \int_0^{\beta}d\lambda \langle 
\{x_S(-i\hbar \lambda), p_S(-i\hbar \lambda) \} p_H(t) \rangle_{\rm eq} ,
\end{equation}
where we have used $\langle p(t) \rangle_{\rm eq}=0$.
The contribution of the third term on the right hand side of (\ref{Omega}) is
\begin{eqnarray}\label{GK3}
\langle p \rangle_{\rm SS}^{(3)}
&=&\frac{i}{2\hbar m}\int_0^{\infty}dt \int_0^{\infty}d\tau 
\LARGE\{\tilde{D}_-(\tau)\int_0^{\beta}d\lambda \langle p_S(-i\hbar\lambda)x_S(-\tilde{\tau}) p_H(t)\rangle_{\rm eq} \nonumber\\
& & {~}
-\tilde{D}_+(\tau) \int_0^{\beta}d\lambda \langle p_S(-i\hbar\lambda)x_S(-\tau) p_H(t)\rangle_{\rm eq} \LARGE\}
.
\end{eqnarray}
The contribution of the fourth term on the right hand side of (\ref{Omega}) is
\begin{equation}\label{GK4}
\langle p \rangle_{\rm SS}^{(4)}
=\frac{1}{2\hbar^2}\int_0^{\infty}d\tau\int_0^{\infty}dt \left( 
\tilde{D}_+(\tau)\langle [x,x_S(-\tau)]p_H(t)\rangle_{\rm eq} 
-\tilde{D}_-(\tau)\langle [x,x_S(-\tilde{\tau})]p_H(t)\rangle_{\rm eq}
\right).
\end{equation}
Thus, we obtain the generalized Green formula
\begin{equation}\label{g-Green-Kubo}
\langle p \rangle_{\rm SS}=\sum_{i=1}^4\langle p \rangle_{\rm SS}^{(i)}.
\end{equation}
It is obvious that three terms 
$\sum_{i=2}^4 \langle p \rangle_{\rm SS}^{(i)}$ 
represents
the nonlinear correction to the conventional Green-Kubo formula, in which the effect of the external force appears through
the time evolution of $p_H(t)$.
We also note that $\langle p \rangle_{\rm SS}^{(2)}$ and $\langle p \rangle_{\rm SS}^{(3)}$ essentially come from
non-dissipative parts, though $p_H(t)$ should involve dissipative effects.
So far, there is no approximation once we start from the basic equations presented in section 2.

It should be noted that the current $p_H(t)$ in Eqs. (\ref{GK1})-(\ref{GK4}) can be replaced by $\Delta p_H(t)\equiv p_H(t)-p_H(\infty)$
if $p_H(\infty)$ is finite because of Eq.(\ref{<Theta>=0}).

\section{Simple example}

In the previous section, we have present formal representations of generalized Green-Kubo formulae, 
but such formal expressions might be insufficient to demonstrate its relevancy. 
In this section, we demonstrate what the result is in the case of $V(x)=0$.
It should be noted that the model is exactly solvable for the harmonic potential but such a case there is no steady current of
the particle because of the trap of the particle in the potential.

If we assume $V(x)=0$, 
$x_S(-\tau)$, $x_S(-i\hbar\lambda)$ and $p_S(-i\hbar\lambda)$ are respectively written as 
\begin{eqnarray}\label{x(t)_hp}
x_S(-\tau)
&=& x+\frac{p}{m}\tau , \\
x_S(-i\hbar\lambda)
&=&x -i\hbar\lambda \frac{p}{m} ,
\label{x_s}
\\
p_S(-i\hbar \lambda)&=& p .
\label{p_s}
\end{eqnarray}

From Eq. (\ref{p_s}) we immediately obtain
\begin{equation}\label{int_ps}
\int_0^{\beta}d\lambda p_S(-i\hbar\lambda)
=\beta p.
\end{equation}
With the aid pf Eqs.(\ref{GK1}) and (\ref{p_H(t).real}), thus, we obtain
\begin{eqnarray}\label{p_ss:first}
\langle p \rangle_{\rm SS}^{(1)}
&=& \frac{F_{\rm ex}}{m}\int_0^{\infty}dt\int_0^{\beta}d\lambda \langle p^2 \rangle_{\rm eq} e^{-2\gamma t}
= \frac{F_{\rm ex}}{2\gamma},
\end{eqnarray} 
where we have used $\langle p^2 \rangle_{\rm eq}=m kT$.
This is the result from the conventional Green-Kubo formula.
We also need to stress that this result is identical to the exact solution of $p_H(t\to \infty)$ in Eq. (\ref{p_H(t).real})
without any statistical average. 
Thus, we expect that the contributions from Eqs. (\ref{GK2})-(\ref{GK4}) are zero in this simple example.
Indeed, it is quite easy to prove the above statement.

Let us evaluate the contribution of Eq.(\ref{GK2}).
From (\ref{x_s}) and (\ref{p_s}), we obtain
\begin{equation}
\{ x_S(-i\hbar\lambda), p_S(-i\hbar\lambda)\}
=xp+px-2\frac{i\hbar\lambda}{m}p^2 .
\end{equation}
Therefore, we  immediately obtain
\begin{equation}
\langle \{ x_S(-i\hbar\lambda),p_S(-i\hbar\lambda)\} p_H(t)\rangle_{\rm eq}=0,
\end{equation}
where we have used  $\langle \{x_S(-\tau),p_S(-i\hbar\lambda) \} \rangle_{\rm eq}=0$ and
function containing odd powers of $x$ or $p$ becomes traceless. Here we note
$p_H(t)$ is given by 
Eq. (\ref{p_H(t).real}).

Let us evaluate the integral terms on the right hand side of (\ref{B3.390}).
By using (\ref{x(t)_hp}) we can write
\begin{equation}\label{1018_1}
[x,\{x_S(-\tau),\rho_{\rm eq}\}]
= [x, \{x,\rho_{\rm eq}\}]+\frac{\tau}{m}[x,\{p,\rho_{\rm eq}\}] .
\end{equation}
Therefore, we directly obtain
\begin{equation}\label{D=0}
{\rm tr}_S\{[x,\{x_S(-\tau),\rho_{\rm eq}\}] p_H(t)\}=0 ,
\end{equation}
where we have used ${\rm tr}_S\{[x,\{x_S(-\tau),\rho_{\rm eq}\}]=0$ and
Eq. (\ref{p_H(t).real}).

Similarly, from
\begin{equation}\label{1020_1}
[x,[x(-\tau),\rho_{\rm eq}]]= [x,[x,\rho_{\rm eq}]]+\frac{\tau}{m}[x,[p,\rho_{\rm eq}]]
\end{equation}
we immediately obtain
\begin{equation}\label{D_1=0}
{\rm tr}_S\{[x,[x_S(-\tau),\rho_{\rm eq}]] p_H(t)\}=0 ,
\end{equation}
where we have used ${\rm tr}_S\{[x,[x_S(-\tau),\rho_{\rm eq}]]=0$ and Eq. (\ref{p_H(t).real}).

Therefore, the generalized Green-Kubo formula for the motion without potential is reduced to the result obtained by
 conventional Green-Kubo formula as
\begin{eqnarray}
\langle p \rangle_{\rm SS}
&=&
\frac{F_{\rm ex}}{2\gamma}
 .
\label{last-harmonic}\end{eqnarray}
To know the nonlinear contributions in Eqs. (\ref{GK2})-(\ref{GK4}) explicitly,
 we need to introduce nonlinear effects of potential or the interaction between particles.

\section{Discussion and Conclusion}

We have obtained the generalized Green-Kubo formula. The final expression should be
nearly equal to (\ref{GK1})-(\ref{GK4}). We also verify the validity of our formulation
in the simplest case for a free Brownian particle under the external force $F_{\rm ex}$,
where the result is obtained from the conventional Green-Kubo formula.  

There are couple of unsolved questions to be answered.
(i) The generalized Green-Kubo formula is not directly related to that we have obtained for classical systems\cite{chong}, 
where $\Theta$ is sum of $\beta \langle \dot{H}_S \rangle_{\rm eq}-\langle \Lambda \rangle_{\rm eq}$, 
where $\langle \Lambda \rangle_{\rm eq}$ is the classical phase volume contraction. 
(ii) Saito\cite{saito08} found that $\langle \dot{H}_S\rangle$ (more precisely, quantum version of Rayleigh's dissipation function)
is directly related to quantum version of Harada-Sasa relation.\cite{harada-sasa} However, the connection between my formulation and
quantum Harada-Sasa relation is not clear.
(iii) We believe that it is straightforward to derive the integral fluctuation theorem in this context without using time-reversed path.
This is the next task.
(iv) Closely related to the above, how to understand generalized Onsager-Casmir relation in this context?\cite{saito07}
(v) How to apply this formulation to the case of microscopic time irreversible quantum systems?\cite{buttiker} 
(vi) We have analyzed a case of steady external force, but we should extend the formulation for the case of time-dependent external field. This is indeed necessary to discuss fluctuation-dissipation relation and its violation. 
(vii) We should analyze the case of nonlinear potentials to clarify the correction of conventional Green-Kubo formula.
In this case, we cannot obtain the exact solution.

\section*{Acknowledgment}

The author thanks S.-H. Chong, M. Otsuki, T. Petrosky and K. Saito for fruitful discussions. 
This work was partially supported by Ministry of Education, Culture, Science and Technology (MEXT), Japan
(Nos. 21015016 and 21540384), and by the Global COE program " The Next Generation of Physics, Spun from
Universality and Emergence" from MEXT Japan. The author also thanks the Yukawa International Program for
Quark-Hadron Sciences at Yukawa Institute for Theoretical Physics, Kyoto University.

\appendix


\section{Generalized Kubo's identity}

Kubo used an important identity in his paper.\cite{kubo} This identity can be written as
\begin{equation}\label{kubo1}
[A, e^{-\beta H_S}]=e^{-\beta H_S}\int_0^{\beta}d\lambda e^{\lambda H_S}[H_S,A]e^{-\lambda H_S}
=-i \hbar e^{-\beta H_S} \int_0^{\beta}d\lambda \dot{A}_S(-i\hbar\lambda)  ,
\end{equation} 
where $A$ is any observable.
One can generalize this identity to Fermionic commutation relation or the case including double commutators as
\begin{eqnarray}\label{kubo2}
\{ A, e^{-\beta H_S} \} &=& 
2e^{-\beta H_S} A + e^{-\beta H_S} \int_0^{\beta} d\lambda e^{\lambda H_S} [H_S,A] e^{-\lambda H_S}  ,
\end{eqnarray}
\begin{eqnarray}
[A, e^{-\beta H_S} B_S (-\tau)]
&=& 
e^{-\beta H_S}[A, B_S(-\tau )] 
+ e^{-\beta H_S} \int_0^{\beta} d\lambda e^{\lambda H_S} [H_S,A] e^{-\lambda H_S} B_S(-\tau)
\nonumber\\
&=& e^{-\beta H_S}[A,B_S(-\tau)]
-i\hbar e^{-\beta H_S} \int_0^{\beta} d\lambda \dot{A}_S(-i\hbar\lambda) B_S(-\tau) .
\label{kubo3}
\end{eqnarray}

The derivation of these identities are straightfoward.
First, we derive eq.(\ref{kubo2}). It is easy to confirm
\begin{equation}
\frac{d}{d\beta} e^{\beta H_S} \{A, e^{-\beta H_S} \}=
e^{\beta H_S} H_S \{A,e^{-\beta H_S}\} + e^{\beta H_S}\{A,\frac{d}{d\beta}e^{-\beta H_S} \}
= e^{\beta H_S}[H_S,A]e^{-\beta H_S} .
\end{equation}
On the other hand, it is easy to confirm the identity
\begin{equation}
\frac{d}{d\beta}\int_0^{\beta}d\lambda e^{\lambda H_S}[H_S,A]e^{-\lambda H_S}=e^{\beta H_S}[H_S,A]e^{-\beta H_S}.
\end{equation}
Thus, we readily obtain eq. (\ref{kubo2}) where we have used $e^{\beta H_S}\{A, e^{-\beta H_S}\}\to 2A$ as $\beta\to 0$.
The derivation of eq.(\ref{kubo3}) is almost identical to the above. If we can use 
$(d/d\beta) e^{\beta H_S}[A, e^{-\beta H_S}B_S(-\tau)]=e^{\beta H_S}[H_S,A]e^{-\beta H_S}B_S(-\tau)$ and
$e^{\beta H_S}[A,e^{-\beta H_S}B_S(-\tau)]\to [A,B_S(-\tau)]$ as $\beta\to 0$, we readily obtain eq.(\ref{kubo3}). 

\section{Explicit calculation of $p_H(t)$}

It is possible to obtain the exact solution of $p_H(t)$ if the potential is harmonic (\ref{h-potential}).
The formal solution $p_H(t)$ is written as
\begin{equation}
p_H(t)=e^{i{\cal L}t}p=\sum_{n=0}^{\infty}\frac{t^n}{n!}(i{\cal L})^n p ,
\label{p_H1}
\end{equation}
where $i{\cal L}$ is given by (\ref{iL}).
Thus, the most important process to obtain $p_H(t)$ is to obtain $i{\cal L}p$ which consists of the five terms as
$ i {\cal L}=\sum_{i=1}^5 i {\cal L}_i$.
The first of $i{\cal L}p$ is
\begin{equation}\label{p_H2}
i {\cal L}_1p\equiv \frac{i}{\hbar}[H_S,p]=0.
\end{equation}
The second term of $i {\cal L}p$ is
\begin{equation}\label{p_H3}
i {\cal L}_2p\equiv \frac{i}{\hbar}[H_c,p]=-2{\cal K} x .
\end{equation}
The third term is given by
\begin{equation}\label{p_H4}
i{\cal L}_3p \equiv -\frac{i F_{\rm ex}}{\hbar}[x,p]=F_{\rm ex} .
\end{equation}
The fourth term is the most complicated, which is given by
\begin{eqnarray}
i {\cal L}_4 p &\equiv &
-\frac{i}{2\hbar^2}\int_0^{\infty}d\tau D(\tau)\{x_S(-\tau),[x,p]\}
=\frac{1}{\hbar}\int_0^{\infty}d\tau D(\tau)x_S(-\tau)
\nonumber\\
&=& \frac{x}{\hbar}\int_0^{\infty}d\tau D(\tau)
-\frac{p}{m}\int_0^{\infty}d\tau D(\tau) \tau
\nonumber\\
&=& -2\gamma p+ 2{\cal K} x ,
\label{p_H5}
\end{eqnarray} 
where we have used (\ref{x(t)_hp}) for the second equality,  and
\begin{eqnarray}\label{int_D}
\int_0^{\infty}d\tau D(\tau)&=& 2\hbar \lim_{\epsilon\to \infty}\int_0^{\infty}d\omega \int_0^{\infty}d\tau J(\omega)e^{-\epsilon \tau}
\nonumber\\
&=& 2\hbar \lim_{\epsilon\to 0} \int_0^{\infty}d\omega \frac{\omega J(\omega)}{\omega^2+\epsilon^2}
=2\hbar {\cal K} . 
\end{eqnarray}
We should emphasize that the contribution of counter Hamilitonian (\ref{p_H3}) 
is cancelled from the contribution of Eq.(\ref{int_D}).
The five contribution to $i{\cal L}p$ is given by
\begin{equation}\label{p_H6}
i {\cal L}_5 p\equiv -\frac{1}{2\hbar^2}\int_0^{\infty}d\tau D_1(\tau)[x_S(-\tau),[x,p]]
=0 .
\end{equation}
Thus, from Eqs.(\ref{p_H2})-(\ref{p_H6}) we obtain
\begin{equation}\label{iLp}
i {\cal L}p= - 2\gamma p+ F_{\rm ex} ,
\end{equation}
or equivalently
\begin{equation}\label{tilde_p}
i {\cal L}\Delta p=-2\gamma \Delta p ,
\end{equation}
where 
\begin{equation}\label{Delta_p}
\Delta p\equiv p-\frac{F_{\rm ex}}{2\gamma)}.
\end{equation}
 Thus, we obtain $ (i{\cal L})^n\Delta p=(-2\gamma)^n \Delta p$.

This leads to
\begin{eqnarray}\label{p_H(t).real}
p_H(t)&=& 
\sum_{n=0}^{\infty}\frac{t^n}{n!}(i{\cal L})^n p =\frac{F_{\rm ex}}{2\gamma}+\left(p-\frac{F_{\rm ex}}{2\gamma}\right)e^{-2\gamma t}.
\end{eqnarray}

\end{document}